\documentclass[%
 reprint,
showpacs,preprintnumbers,
 amsmath,amssymb,
 aps,
]{revtex4-1}
\usepackage{graphicx}
\usepackage{dcolumn}
\usepackage{bm}
\usepackage{epstopdf}
\usepackage{amsbsy}
\usepackage{amsmath}
\usepackage{amssymb}
\usepackage[separate-uncertainty = true,multi-part-units=single]{siunitx}
\usepackage{commath}
\usepackage{graphicx}
\usepackage{color}
\usepackage[final]{changes}
\usepackage[utf8]{inputenc}
\usepackage[ngerman,english]{babel}
\usepackage{mathtools}

\newcommand{\EQ}[3]{
  \begin{equation}
    \label{#1}
    #2
    \;#3
  \end{equation}
}
\newcommand{\ER}[1]{Eq.~(\ref{#1})}
\newcommand*\diff{\mathop{}\!\mathrm{d}}
\newcommand*\Diff[1]{\mathop{}\!\mathrm{d^#1}}

\begin{document}

\preprint{APS/123-QED}

\title{Examining nonextensive statistics in relativistic heavy-ion collisions}

\author{A. Simon and G. Wolschin}%
 \email{g.wolschin@thphys.uni-heidelberg.de}
\affiliation{%
 Institut f{\"ur} Theoretische 
Physik
der Universit{\"a}t Heidelberg, Philosophenweg 16, D-69120 Heidelberg, Germany, EU\\
}%


\date{\today}

\begin{abstract}
We show in detailed numerical solutions of the nonlinear Fokker-Planck equation (FPE) 
which has been associated with nonextensive $q$-statistics that the available data
on rapidity distributions for stopping in relativistic heavy-ion collisions 
cannot be reproduced with any permitted value of the nonextensivity parameter ($1 < q < 1.5$). 
This casts doubt on the nonextensivity concept that is
widely used in relativistic heavy-ion physics.

\end{abstract}

\maketitle
\section{Introduction}
Nonextensive statistics proposes an extension of Boltzmann statistics through the concept of a non-additive $q$-entropy. 
It has been used in a nonlinear Fokker-Planck equation (FPE) for rapidity distributions, 
and applied to calculate rapidity and transverse momentum distributions for produced and stopped charged particles in relativistic heavy-ion collisions. 
In the corresponding experiments, the measured charged-hadron rapidity distributions are found to be very broad compared to 
thermal model predictions \cite{sta96}, and the discrepancy increases strongly with energy. 
This finding, as well as correspondingly broad net-proton (proton minus antiproton, or stopping) distributions \cite{app99,bea04}, indicates thermal diffusion plus collective expansion. 

Both effects may be accounted for phenomenologically in a linear diffusion model \cite{gw99} with expansion, 
or else in abundant hydrodynamic approaches (see e.g.~\cite{hesne13} for a review). It has, however, been stipulated that the so-called
nonextensive $q$-statistics  as proposed by Tsallis et al.~\cite{tsa96} can simultaneously account for thermal and collective effects merely through a suitable choice of $q$ \cite{lav02,alb08,alb09}.

With values of $1<q<1.5$, the Fokker-Planck equation that has been used to model rapidity distributions \cite{gw99} becomes nonlinear, it has an
exponent $(2-q)$  in the diffusion term \cite{tsa96,lav02,alb08,alb09}. This is supposed to account for long-range forces that cause collective expansion, and is considered
to be a fundamental property of the system like the temperature $T$. It goes along with a modified definition of the system's entropy \cite{tsa96} which is, however, controversial on fundamental grounds \cite{nau03,bana06}. This approach is quoted in \cite{lav02} as having the additional reward that the Einstein relation between drift and diffusion coefficient -- that is valid in \replaced{the theory of}{linear theory such as the one for} Brownian motion -- could be maintained.

Indeed it has been claimed \added{in} \cite{lav02,alb08} that such a procedure provides fits to stopping data in PbPb and AuAu collisions at energies reached at the Super Proton Synchrotron (SPS)
and the Relativistic Heavy Ion Collider (RHIC) with values $1<q<1.5$. (No stopping distributions will be available in the foreseeable future from the Large Hadron Collider (LHC) due to the lack of a suitable forward spectrometer). 
However, the solutions of the nonlinear FPE in these calculations are obtained by starting from the stationary solution as proposed in \cite{tsa96}, and
then solving the problem for time-dependent temperatures $T(t)$ and mean values of the rapidity $y_m(t)$. 

It is the aim of this work to solve the nonlinear FPE directly with realistic physical initial conditions. We shall use several independent numerical schemes, but without a pre-determined form of the solutions -- such as taking the form of the stationary solution as a basis for the time-dependent case as had been done in \cite{lav02} --, 
and then try to fit the measured stopping distributions at SPS and RHIC energies with a value of $q>1$.

Some relevant model ingredients for the linear and nonlinear cases are summarized in the next two sections. The numerical calculations are prepared and tested in the subsequent section. In particular, their implementation is compared to the exact solution of the linear case, and to a specific exact solution of the nonlinear case found by Borland et al.~\cite{bor99}. The latter
provides a precise test of the numerical methods, but it is not useful for a solution of the physical problem with $\delta$-function initial conditions at the beam rapidities.
We also compare three completely independent solution schemes for the nonlinear problem with each other.

In the final section, we apply the numerical solution of the nonlinear FPE to the calculation of net-proton distributions in PbPb and AuAu collisions at SPS and RHIC energies,
and show that it is not possible to fit the data using solutions of the nonlinear FPE with values of the nonextensivity coefficient $1<q<1.5$. Instead we fit the data in the linear model  \added{\cite{fgw17}} with an adjusted diffusion coefficient to account for both, nonequilibrium thermal broadening, and collective expansion. The results are then briefly summarized.

\section{Basic considerations}
In relativistic heavy-ion collisions, the relevant observable in stopping and particle production is the Lorentz-invariant cross section
\EQ{eq:3e}{E \frac{\Diff3 N}{\diff p^3} = \frac{\Diff2 N}{2 \pi p_\perp \diff p_\perp \diff y} = \frac{\Diff2 N}{2 \pi m_\perp \diff m_\perp \diff y}}{}
with the energy $E = m_\perp\cosh(y)$, the transverse momentum $p_\perp = \sqrt{p_{x}^2+p_{y}^2}$, the transverse mass $m_\perp = \sqrt{m^2+p_\perp^2}$,  and 
the rapidity $y$. In this work\added{,} we concentrate on rapidity distributions of protons minus produced antiprotons, which are indicative of the stopping process as described 
phenomenologically in a relativistic diffusion model (RDM) \added{\cite{gw99,fgw17}}, or in a QCD-based approach \cite{mtw09}. The rapidity distribution is then obtained by integrating over the transverse mass 
\EQ{eq3:f}{
  \frac{\diff N}{\diff y}(y,t) = C \int\; m_\perp E \frac{\Diff3 N}{\diff p^3} \diff m_\perp
}{,}
with a normalization constant $C$ that depends on the number of participants at a given centrality. The experimentally observable distribution \replaced{$\text{d}N/\text{d}y$}{$dN/dy$} is calculated for the
freeze-out time, $t=\tau_\text{f}$\,. The latter can be identified with the interaction time $t=\tau_\text{int}$ of \cite{gw99,fgw17}: the time during which the system interacts strongly.

We rely on Boltzmann-Gibbs statistics and hence, adopt  the Maxwell-J{\"u}ttner distribution
as the thermodynamic equilibrium distribution for $t\rightarrow \infty$ 
\begin{eqnarray}
\label{eq:2b}
  E \frac{\text{d}^3N}{\text{d}p^3} \Bigr|_\text{eq}\propto E \exp\left(-E/T\right) \qquad\qquad\nonumber \\\\ 
  \equiv m_\perp \cosh\left(y\right) \exp\left(-m_\perp \cosh(y) / T\right).\nonumber
\end{eqnarray}
In thermodynamics\added{,} one makes the distinction between extensive and intensive properties.
Intensive properties do not depend on the size of the system or the amount of mass inside the system.
These are for example the temperature or the mass density.
Extensive properties on the other hand are proportional to the mass and increase as the size of the system increases.
Typical examples are the volume and the mass itself.

In statistical physics, the entropy is also extensive:
The Boltzmann-Gibbs\deleted{'} definition of the entropy is $S = -k_{B} \sum_{i=1}^{\Omega} p_{i}\;\ln(p_{i})$, where $p_i$ equals the probability of the system to be in the microstate $i$.
In the case of equal probabilities and a total number of states $\Omega$ it follows that $p_i = p = \frac{1}{\Omega}$\added{,} and (with $k_B\equiv 1$)

\EQ{eq:1}{\begin{split}
    S & = -\sum_{i=1}^\Omega \frac{1}{\Omega}\;\ln\left(\frac{1}{\Omega}\right)
    = - \sum_{i=1}^\Omega \frac{1}{\Omega}(0-\ln(\Omega)) \\
    & =  \ln(\Omega),
  \end{split}}
{}
which is the well-known expression for the entropy.
To show its extensivity\added{, one} take\added{s} two systems $A$ and $B$ which do not interact.
The number of available microstates in the combined system is equal to the product of the ones in the individual systems as they do not interact,
\EQ{eq:2}{\Omega(A+B) = \Omega(A)\;\Omega(B)}{.}
Inserting this into the definition of entropy\added{,} one gets
\EQ{eq:3}{
\begin{split}
  S(A + B) & = \ln(\Omega(A + B))  \\
    & = S(A) + S(B)\,.
\end{split}}{}
Hence, the Boltzmann-Gibbs entropy is an extensive property of the system.

Although classical thermodynamics is a very successful theory, discrepancies with respect to data can arise. This is 
particularly relevant in the case of nonequilibrium systems, such as 
relativistic heavy-ion collisions. However, statistical mechanics 
is then still built upon the principle that the information $I$ is minimized with 
constraints that are appropriate for the given physical situation, and the
entropy is uniquely defined as $S=-k_B I$.

Nevertheless, different concepts of entropy have been developed
for nonequilibrium systems. In particular, Tsallis has proposed to resort to 
nonextensive statistics \cite{tsa88,tsa96} where the entropy does not fulfill Eq.\,(\ref{eq:3}) 
but is instead given by
\EQ{eq:3ah}{
  \begin{split}
    S_q &= \langle \ln_q \frac{1}{p_i} \rangle = \sum p_i \ln_q \frac{1}{p_i} \\
    &= \frac{\sum p_i - \sum p_i^q}{q-1} = \frac{1 - \sum p_i^q}{q-1}
  \end{split}
}{}
with the entropic index $q\in \mathbb{R}$.
Here\added{,} the logarithm which causes the additivity of the entropy has been replaced by the non-additive $q$-logarithm $\ln_q(x)$ such that
$S_q(A+B)=S_q(A)+S_q(B)+(1-q) S_q(A)S_q(B)$, and $q$ measures the degree of nonextensivity. 
The inverse of the $q$-logarithm is the $q$-exponential $e^x_q$
that solves the differential equation d$y/\text{d}x=y^q$ through
\EQ{eq:3ac}{
  y = \left[1+(1-q)\,x\,\right]^{1/(1-q)} \equiv e^x_q}
{.}
In the limit $q \rightarrow 1$, $S_q$ is equal to $S$ because
\EQ{eq:3ai}
{
  \begin{split}
    p_i^q &= e^{q\ln(p_i)} = e^{(q-1)\ln(p_i) +\ln(p_i)} \\
    &= e^{(q-1)\ln(p_i)}p_i = p_i\left[1+(q-1)\ln(p_i)\right]
 + O(\|q-1\|^2)
  \end{split}
}{}
provided the last term in Eq.\,(\ref{eq:3ai}) is neglected,
\EQ{eq:3aj}{
  \begin{split}
    S_{q\rightarrow 1} &= \frac{1-\sum p_i \left[1+(q-1)\ln(p_i)\right]}{q-1} \\
    &= \frac{ 1 - \sum p_i + (q-1) \sum p_i \ln p_i}{q-1} \\
    &= \sum p_i \ln p_i = S\;.
  \end{split}
}{}
There is, however, no clearly defined physical process that would warrant a generalization from $S$ to $S_q$, and no theory available to calculate
the nonextensivity exponent $q$ from first principles. It can still successfully be used as an additional fit parameter, in particular for $p_\perp$-distributions
in $pp$ and $AA$ collisions at relativistic energies which show a transition from exponential to power-law behaviour that the $e^x_q$-function properly describes with
$q \in (1,1.5)$.
From a more fundamental point of view, the approach is controversial \cite{nau03,bana06}. In this work, we test its applicability to rapidity distributions in relativistic heavy-ion collisions.

\section{Fokker-Planck equation}

The general form of the linear Fokker-Planck equation (FPE) is \cite{risk96}
\EQ{eq:3ba}{
  \pd{}{t}W(y,t) = -\pd{}{y}[J(y,t)W(y,t)] + \pd[2]{}{y}[D(y,t)W(y,t)]
}\\
 where $J$ is called the drift coefficient and $D$ is the diffusion coefficient. Here we denote the 
 independent variable as $y$ because it will later considered to be the rapidity.
The FPE can also be written in the form of a continuity equation for the probability distribution $W$ as
\EQ{eq:3bb}{
  \pd{}{t}W+\pd{}{y}j = 0
}{,}
with \added{$j(y,t) = \left[J(y,t)-\pd{}{y}D(y,t)\right]W$}, which is interpreted as a probability current~\cite{risk96}.
Even for coefficients $J$ and $D$ that are not time dependent it is generally difficult\replaced{ --}{,} if not impossible\added{ --} to find analytical solutions.
Two important \replaced{analytically solvable examples are}{examples where solutions do exist are for} $J(y,t) = 0$\replaced{,}{ and} $D(y,t) = D$ (Wiener process) \replaced{and}{or} $J(y,t)=-\alpha y$\replaced{,}{ and} $D(y,t) = D$ (Uhlenbeck-Ornstein process \added{\cite{uhl30}}).
For more complicated problems, \deleted{especially nonlinear ones,} numerical methods are employed.

In the relativistic diffusion model\added{,} the time evolution of the rapidity spectra has been modeled by a FPE.
At first \replaced{an Uhlenbeck-Ornstein (UO) ansatz}{a linear drift ansatz} has been tested\added{ in}Ref.~\cite{gw99}.
The stationary solution in such a case is determined as
\EQ{eq:3bc}{
  \pd{}{y}\left[\alpha yW+D\pd{}{y}W\right]=0 \implies \pd{W}{y} \propto -yW + C}
{.}
$C$ has to be equal to zero because otherwise $W < 0$,
{\EQ{eq:3bd}{
  \begin{split}
    \int \frac{1}{W} \dif W &\propto \int -y \dif y  \implies \ln W \propto -\frac{1}{2}y^2+\text{const }\\
    & \implies W \propto e^{-\frac{1}{2}y^2}\,.
  \end{split}}{}
This does not correspond to the equilibrium distribution from \ER{eq:2b} and therefore another drift term is needed.
We see from the above calculation that a stationary solution $ W \propto e^{-V(y)}$ results from a drift term $V'(y)$.
With the drift 
\begin{equation}
J(y)=-A\sinh(y)
\label{drift}
\end{equation}
one gets the desired stationary solution \added{\cite{lav02,fgw17} with}\deleted{. Using a special form of the so-called fluctuation-dissipation theorem -- similar to the relation $D=bT$ in Brownian motion
with the mobility $b$ --, namely \cite{lav02,fgw17}}
\begin{equation}
A=\frac{m_\perp D}{T}\,,
\label{fdt}
\end{equation}
\added{which can be interpreted as a fluctuation-dissipation relation similar to one known from Brownian motion, $D=bT$ 
with the mobility $b$. Hence,}  
the dissipation as described by the amplitude of the drift term can be related to the diffusion coefficient that is responsible for the fluctuations.

This particular \added{sinh-}drift term has also been investigated in Ref.~\cite{fgw17} and the result was -- as in the simple \replaced{UO model}{model with linear drift} \cite{wols99} -- that the fluctuation-dissipation relation is violated\deleted{ in case of a linear diffusion term}: The diffusion is too small to account for the experimental data.
\par
The canonical interpretation of this result is that collective expansion occurs in the quark-gluon-plasma phase and enhances the width.
One way to match the observation is to\deleted{ simply} increase the diffusion coefficient, attributing the effect to collective expansion~\cite{wols99,fgw17}.
\replaced{Indeed a general form of the fluctuation-dissipation theorem has been used in relativistic hydrodynamic calculations that describe systems exhibiting longitudinal collective expansion \cite{kap11}.}{Indeed it has been shown in relativistic hydrodynamics that the fluctuation-dissipation theorem itself can be generalized such that it is consistent with systems exhibiting longitudinal collective expansion \cite{kap11}.}

Within the Fokker-Planck framework, another possibility is to change the underlying equation in order to account for the \replaced{`}{'}anomalous' diffusion \cite{lav02,alb08,alb09}.
In the latter approach which we want to test here, one extends the model to a nonlinear FPE \deleted{(NLFPE)}
 
  \begin{eqnarray}
  \label{eq:3be}
    \pd{}{t}W(y,t)^{\mu} = -\pd{}{y}[J(y,t)W(y,t)^{\mu}] \qquad\qquad\qquad\\
   + \pd[2]{}{y}[D(y,t)W(y,t)^{\nu}]\,.\nonumber
\end{eqnarray}

Analytical solution strategies for this equation in case of $\nu \ne 1, \mu \ne 1$ are not readily
available. However, one can connect Eq.\,(\ref{eq:3be}) with the nonextensive entropy Eq.\,(\ref{eq:3ah}).
Indeed, Tsallis and Bukman have shown in~\cite{tsa96} that the result of maximizing the entropic form

\EQ{eq:3bf}
{
  S_q[p] = \frac{1 - \int \dif u \left[p(u)\right]^q}{q-1}
}{,}
leads to the function
\EQ{eq:3bg}
{
 p_q(y,t) = \frac{\left\{ 1-\beta(t)(1-q) \left[ y-y_m(t) \right]^2 \right\}^{\mathrlap{1/(1-q)}}}{Z_q(t)}
}
{.}\\

When assuming a \replaced{drift term $J(y,t)=-\alpha y$ and a constant diffusion coefficient $D(y,t)=D$, 
the function $ p_q(y,t) \equiv W(y,t)$ solves the partial differential equation Eq.~(\ref{eq:3be})}{linear drift term $J(y,t)$ in {Eq.~(\ref{eq:3be}),}
this function solves the partial differential equation (PDE)} with additional conditions on $\beta(t)$, $y_m(t)$ and $Z_q(t)$\replaced{. O}{, and o}ne can identify $q$ from the entropic form with the exponents $\mu$ and $\nu$ of \deleted{the NLFPE}\ER{eq:3be} as $q = 1 + \mu - \nu$ \cite{tsa96}.

This identification is actually only  justified in the case of \added{the above} linear drift, which is not the one we will use because the Boltzmann equilibrium form requires a $\sinh$-drift.
It was also shown in Ref.\,\cite{tsa96} that in order to conserve the norm, $\mu = 1$ is required, and since we model a probability distribution we set $\mu$ to one, such that
the exponent of the diffusion term becomes $\nu=2-q$.
\replaced{Rewriting}{Differentiating} the diffusion term\deleted{ once} as
\EQ{eq:3bh}{
  \pd[2]{}{y}\left[DW^{2-q}\right] = \pd[2]{}{y}\left[(DW^{1-q})\,{W}\right]
}{,}
 we can view the nonlinearity in the exponent as ordinary diffusion extended by a nonlinear diffusion coefficient, namely
$D' = D W^{1-q}$.
It is visualized in Fig.\,\ref{fig1}.
\begin{figure}
\begin{center}
\includegraphics[width=8.6cm]{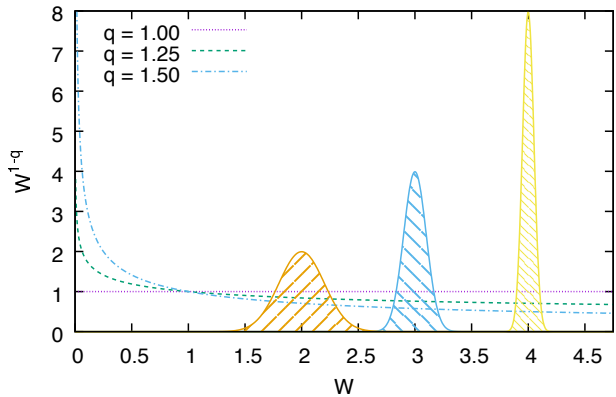}
\caption{\label{fig1}(Color online) Comparison of the diffusion nonlinearity $W^{1-q}$ for different $q$ values.
  The coefficient is variable for $q > 1$ and depends on the size of the probability distribution itself.
 \deleted{To get an idea about the size of (normalized) probability distributions we also plotted }Gauss curves with $\sigma=0.2, 0.1, 0.05$
  \added{ are plotted as a reference for the size of the normalized probability distribution function}.
  \deleted{The height of the Gauss curve then corresponds to $W$.} }
\end{center}
\end{figure}
\begin{figure}
	\centering
\includegraphics[width=8.6cm]{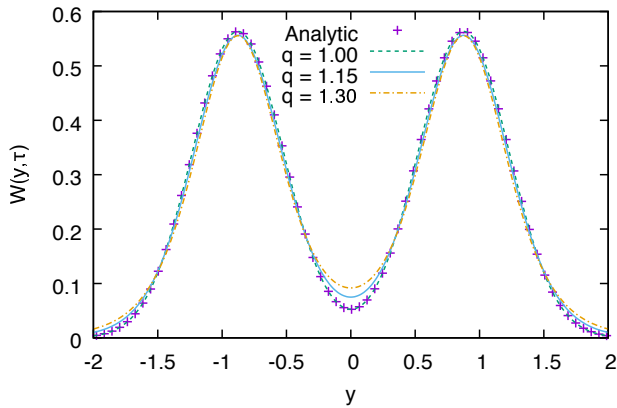}
\caption{\label{fig2}(Color online) Comparison of the analytic \replaced{Uhlenbeck-Ornstein (UO)}{linear drift} model (crosses) and the corresponding numerical solution ($q=1$, dashed curve). The numerical solutions for two different values of $q$ are also shown (solid for $q=1.15$, dash-dotted for $q=1.3$).
The parameters are $\gamma = 0.137$, $y_0 = 2.9$, $\sigma = 0.1$, $\tau = 1.2$.}
\end{figure}
\begin{figure}
	\centering
\includegraphics[width=8.6cm]{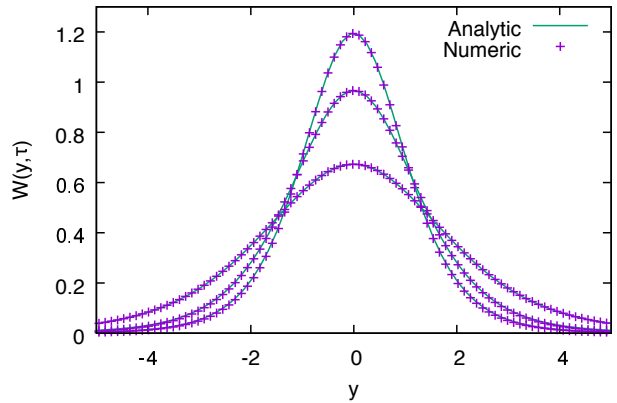}
\caption{\label{fig3}(Color online) Comparison of the analytical (solid curves) and numerical (crosses) solutions for $q = 1.3$ at different dimensionless times $\tau = $ 0.2, 0.5, 0.8.
(Top to bottom at $y=0$.}
\end{figure}
For function values of less than one, the diffusion is increased and for values larger than one it is suppressed.
This leads to thinner peaks and faster diffusion in the tails.

The width that we used for the simulation ($\sigma = 0.1$), represented by the middle curve in Fig.\,\ref{fig1}, peaks at $y=4$.
 This results in diffusion coefficients of \num{0.5} and \num{0.75}, depending on the \added{value of} $q$.
In the tails\added{,} the diffusion amplification peaks at a factor of \num{5} or \num{6} before the distribution functions get negligibly small.

\section{Numerical calculations}  
 
\subsection{General procedure}

To arrive at a usable form for the computer\added{,} we transform the equation for $W(y,t)$ into its dimensionless version for $f(y,t)$ by introducing a new timescale $t_c$, resulting in the
dimensionless time variable $\tau = t / t_c$.
It follows that $\pd{}{t} = \pd{}{\tau} t_c^{-1}$ and further
\EQ{eq:4}{
  \pd{f}{\tau} = t_c\;A\;\pd{}{y}\left[\sinh(y)\;f(y,t)\right] + t_c\;D\;\pd[2]{}{y}\left[f(y,t)^{2-q}\right]
}{.}

Since \replaced{$A = m_\perp D/T$}{$A = \frac{m_\perp D}{T}$} we set \replaced{$t_c =T/(m_\perp D) = A^{-1}$}{$t_c = \frac{T}{m_\perp D} = A^{-1}$}.
The result is the dimensionless \ER{eq:4aa} depending only on the ratio \replaced{$\gamma = T/m_\perp$}{$\gamma = \frac{T}{m_\perp}$} of temperature $T$ and transverse mass $m_\perp$ which is a measure of the strength of the diffusion,
\EQ{eq:4aa}{
    \pd{f}{\tau} = \pd{}{y}\left[\sinh(y)\;f(y,t)\right] + \gamma\;\pd[2]{}{y}\left[f(y,t)^{2-q}\right]
}
{.}
To get physical values for the drift and diffusion coefficients\added{,} one has to specify a time scale (or the other way round).
Considering that it is only the drift term that is responsible for determining the peak position\added{,} we are free to chose the time $\tau$ such that the peak position of the experimental data is reproduced.
This leaves as free parameters the diffusion strength $\gamma$ and the nonextensivity parameter $q$.

We calculate the solution using two different methods\deleted{,} in order to gain insight about the accuracy.
The more straightforward one was using \textsc{matlab}'s integration routines for solving parabolic-elliptic PDEs.
The second, more elaborate method, was implementing it in a finite-element-method framework (FEM) (DUNE~\cite{ba10} and FEniCS~\cite{aln15}).

To make use of the FEM\added{,} we have to convert our PDE into the so-called weak formulation which reformulates the problem as an integral equation.
This is done by integrating  the \replaced{left-hand-side(LHS)}{lhs} of \ER{eq:4aa} over the whole domain $\Omega \subset \mathbb{R}$ and multiplying it by a test function $g(y)$ that vanishes on the boundary $\partial \Omega$,
\EQ{eq:4b}{
  \int_{\Omega} \dif{y}\;\left\{g(y)\;\pd{}{y}\left[\sinh(y)\;f(y,t) + \gamma\;\pd{}{y}f(y,t)^{2-q}\right]\right\}
}{.}
Integrating this by parts\added{,} we get
\EQ{eq:4c}{\begin{split}
    \left[g(y)\left\{\sinh(y)f(y,t) + \gamma\pd{}{y}f(y,t)^{2-q}\right\}\right]\Biggr\rvert_{\partial \Omega} \\
     - \int_{\Omega} \dif{y}\;\left\{\od{g}{y}\left[\sinh(y)f(y,t) + \gamma\;\pd{}{y}f(y,t)^{2-q}\right]\right\}\;.
\end{split}
}{}
The first line in \ER{eq:4c} vanishes because of $g$ and the second line contains only first derivatives.
To approximate the time derivative in \replaced{\ER{eq:4aa} (LHS)}{\ER{eq:4b} (RHS)}\added{,} we use the Backward-Euler scheme
\EQ{eq:4d}{
  \pd{f(t_n)}{t} = \frac{f(t_n)-f(t_{n-1})}{\Delta t} + O(\|\Delta t^2\|)
}{.}
For both methods the chain rule is used to write $\pd{}{y} f^{2-q}$ as $(2-q) f^{1-q} \pd{}{y} f$.
Because we analyze cases for $q > 0$\added{,} we have to take care of the singularity at $f=0$.
To get around this issue\added{,} we add a small constant to the argument stabilizing the computation: $\left(f+\epsilon\right)^{1-q}$.
In \textsc{matlab} we use the routine \texttt{pdepe} to integrate the equation.
It is suited for parabolic-elliptic problems and we could directly insert the PDE without modifying it.

To compare the simulation to experimental data, we have to insert relevant values for $T$, $m_\perp$ and the initial conditions, most importantly $y_0$.
The value of the beam rapidity $y_0$ is determined by the center-of-mass energy per nucleon pair as $y_0 = \ln(\sqrt{s_{NN}}/m_p)$.
Two Gaussian distributions centered at $\pm y_0$ with a small width $\sigma$ that corresponds to the Fermi motion represent the incoming ions before the collision.
The exact value of $\sigma$ does not have a large effect on the time evolution \cite{fgw17}; here we use a value of \num{0.1}.

For the temperature\added{,} we take the critical value \SI{160}{\mega\electronvolt}  for the transition between hadronic matter and quark-gluon plasma.
The actual freeze-out temperature is smaller ($T$ = \SI{118 \pm 5}{MeV} for PbPb at SPS energies \cite{app99});
overestimating the temperature will increase the diffusion.
For 17.2 GeV PbPb, the transverse mass is taken to be $m_\perp=$ \SI{1.17}{\giga\electronvolt} as the average transverse momentum $p_\perp$ is around \SI{0.7}{\giga\electronvolt}~\cite{app99}.
The dimensionless diffusion strength $\gamma$ is thus \num{0.137}. Corresponding values for 200 GeV AuAu will be given later.
 
 The results are then transformed to a rapidity distribution \cite{fgw17}.
 Rewriting Eq.\,(\ref{eq3:f}) and replacing \replaced{$\text{d}^3N/\text{d}p^3$}{$d^3N/dp^3$} with the computed
distribution $f(y,t)$, we obtain
\EQ{eq:4db}
{
    \od{N}{y}(y,t) = C \int m^2_\perp \cosh(y) f(y,t) \dif m_\perp \\\,.
}{}
Since the transverse mass $m_\perp$ is mainly distributed around $m_p$ \cite{app99},
we introduce an upper integration limit $m^*$ such that the second
moment of $m_\perp$ corresponds to the measured value \cite{app99} at SPS energies, and accordingly at RHIC energies
\begin{equation}
\langle m^2_\perp\rangle=\int^{m^*}_{m_p} m^2_\perp \dif m_\perp \,.
\end{equation}
The rapidity distribution for net protons can then approximately be written as
\begin{equation}
  \od{N}{y}(y,t) \approx \tilde{C} \langle m^2_\perp\rangle \cosh(y) f(y,t)\,.
  \end{equation}
The constant $\tilde{C}$ is chosen such that the total number of particles for $0-5\%$ corresponds to the number
of participant protons in this centrality bin.\\

\begin{figure}
	\centering
\includegraphics[width=8.0cm]{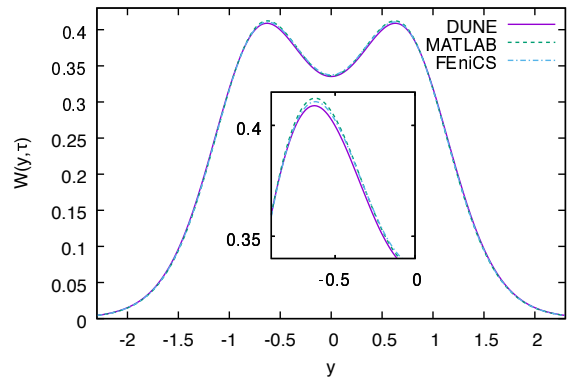}
\caption{\label{fig4}(Color online) Comparison of the three numerical solution methods for $\tau = 1$ and $\gamma = 0.137$.}
\end{figure}
\begin{figure}
\centering
\includegraphics[width=8.0cm]{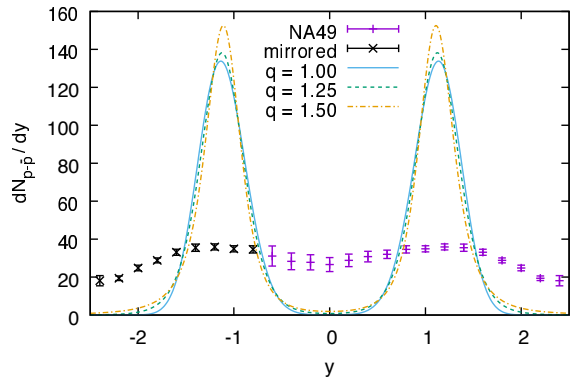}
\caption{(Color online) Numerical solutions of the nonlinear FPE for central PbPb at 17.2 GeV with three different values of $q\in [1,1.5]$, and NA49 data \cite{app99}.} 
\label{fig5}
\end{figure}

\subsection{Tests of the numerical implementation}

In order to check the numerical  implementation\added{,} we compare it to analytically solvable problems.
At first\added{,} we consider the \replaced{UO}{linear drift} model and compare the numerical solution \added{of Eq.\,(\ref{eq:4aa})} for different values of $q$, first for ${q=1}$ (where both should be the same) and then for other values of $q$, 
see Fig.\,\ref{fig2}.
This gives us a first idea about the impact of the non-linearity parameter on the evolution.

The numerical result for $q=1$ is identical with the analytical solution, which validates the numerical method.
By increasing $q$\added{,} the peaks are slightly smeared out, giving an overall flatter shape than before.
This is expected since a larger diffusion coefficient will spread out the profile faster.

As the next \replaced{step,}{problem} we \replaced{consider the problem}{take the one} solved analytically by Borland et al.~in Ref.~\cite{bor99}
\EQ{eq:5a}
{
  \pd{f}{t} = \pd{}{y}\left(y f\right) + \pd[2]{}{y}f^{2-q}
}{.}
\deleted{The results are displayed in Fig.\,\ref{fig3}.}
The solution assumes that the initial condition is functionally equal to the stationary solution, except for time-dependent coefficients.
In particular, both the stationary solution, and the initial conditions are centered at $y=0$, which is essential to obtain the
analytical solution of the time-dependent problem.
\deleted{It is thus not useful for our physical situation with initial conditions that are adapted to the heavy-ion collision case, 
 but we get the possibility to compare the anomalous diffusion to an analytic solution.} 
 
 In the case of a heavy-ion collision, \added{however,}
 the initial distributions are both off-center at the values of the beam rapidities, whereas the stationary solution that is obtained
 for $t\rightarrow\infty$ is centered at rapidity $y=0$ in symmetric systems\replaced{. Hence,}{and hence,} the Borland et al.~analytical solution cannot be used\replaced{:}{.} 
 The time-dependent equation must be solved with the beam rapidities
 $y_\text{beam}=\pm y_0$ defining the initial conditions ($\delta$-functions, or Gaussians with a width that is determined by the
 Fermi motion), and the solution in the heavy-ion case drifts with increasing time towards midrapidity. \added{Although the Borland solution cannot describe our physical situation, it offers the possibility to compare the anomalous diffusion to an analytic solution.}

\begin{figure}
\centering
\includegraphics[width=8.0cm]{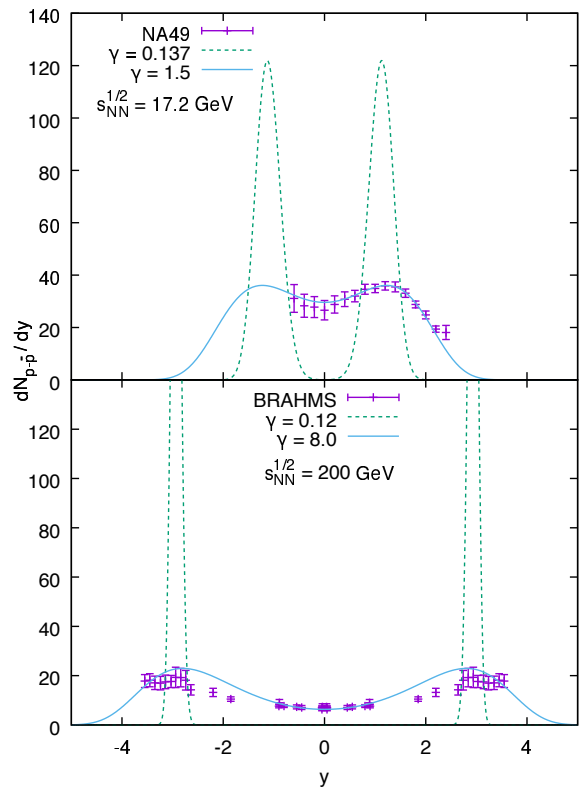}
\caption{(Color online) Comparison of the \replaced{linear ($q=1$)}{linear} model without (dashed) and with (solid) adjusted diffusion term with NA49 data for 0-5\% central PbPb at $\sqrt{s_{NN}} = 17.2$ GeV \cite{app99}, upper frame, and with BRAHMS data for central AuAu at $\sqrt{s_{NN}} = 200$ GeV \cite{bea04}, lower frame. The values of the dimensionless diffusion strengths are $\gamma=0.137$ and $0.12$ from the fluctuation-dissipation relation (see text), whereas $\gamma=1.5$ and $8.0$ are adjusted to the SPS and the RHIC data, respectively, and account also for collective expansion. The values of the freeze-out time have been adjusted in both cases. A numerical solution of the nonlinear diffusion equation with $q>1$ does not fit the data for any value of $\gamma$ and time.}
\label{fig6}
\end{figure}
  
The agreement between analytical and numerical solution \added{(see Fig.\,\ref{fig3})} in the case of initial conditions that are centered at $y=0$ further supports the correctness of the implementation.
We have now three numerical schemes at our disposal to calculate the evolution.
Since they are based upon two different mathematical methods (finite elements and finite differences) it is unlikely that a hypothetical programming error occurred in all of them.
Having this in mind\added{,} we simulated the full PDE using each of the packages and compared the results.
The time evolution of two Gaussian peaks with $y_0 = 2.9$ and $\sigma = 0.1$ at the time $\tau = 1$ is shown in Fig.\,\ref{fig4}.

The relative difference between the solutions using the three numerical schemes is around \SI{1}{\percent} and mostly concentrated at the peaks.
Possible origins of the slight discrepancies are the different step sizes used in each discretisation and the basis functions used in the FEM interpolation.
In any case\added{,} the differences are very small, from which we conclude that the calculations are correct.
Since further data analysis is easiest in \textsc{matlab}, we use it in the following calculations.

\section{Comparison of numerical results and experimental data}

The results of the calculation for different values of $q$ are shown in Fig.\,\ref{fig5} for PbPb at 17.2 GeV.
While a larger $q$ does broaden the distribution, the effect is by far too small to come close to the experimental results.

In order to reproduce the measured data for PbPb\added{,} we have to adopt a diffusion strength of around \num{1.5} while the one predicted by the fluctuation-dissipation relation 
Eq.\,(\ref{fdt}) is around \num{0.137}, the difference being a factor \num{11}, see the upper frame of Fig.\,\ref{fig6}.
As we mentioned earlier and as can be seen in Fig.\,\ref{fig1}, such a large enhancement in the required broadening cannot be compensated by the proposed nonlinearity
due to $q$-statistics. 

The comparison with AuAu stopping data at the maximum energy of 200 GeV reached at RHIC shows that here the discrepancy between the diffusion strength from the
fluctuation-dissipation relation ($\gamma=0.12$) and the one required to fit the data ($\gamma=8$) with an adjusted value of time is even larger, see lower frame of Fig.\,\ref{fig6}.
This means that introducing a nonlinearity into the diffusion term cannot account for the observed rapidity spectra at SPS and RHIC energies.
Since the widths are too narrow, there has to be an additional expansion process that takes place during the reaction that cannot be accounted for by $q$-statistics.
This result is in obvious contrast to the findings of Refs.~\cite{lav02,alb08,alb09}, where an approximate solution of \ER{eq:4aa} had been used.

We have also solved the nonlinear diffusion equation separately for initial conditions centered at $y_\text{beam}=+y_0$, and at $y_\text{beam}=-y_0$ to assess how much the superposition principle is violated in the nonlinear case. Adding the results 
shows that the difference with respect to the full numerical solution remains, however, below $5\%$ at midrapidity.

As the numerical solution of the nonlinear \ER{eq:4aa} does not explain the experimental data, we return to the model with linear diffusion $q = 1$, and the drift term imposed by the stationary solution \cite{fgw17}. 
By fitting experimental data to this linear model\added{,} we can find physical values for the drift and diffusion coefficients in stopping using the
two data sets from  NA49~\cite{app99} at $\sqrt{s_{NN}} = $ \SI{17.2}{\giga\electronvolt} with beam rapidity $y_\text{beam}=\pm 2.91$, and \added{from }BRAHMS~\cite{bea04} at  $\sqrt{s_{NN}} = $ \SI{200}{\giga\electronvolt} with $y_\text{beam}=\pm 5.36$ in central collisions of PbPb and AuAu, respectively.
With a freeze-out time of \added{\SI{8}{fm/\textit{c}}}, we obtain
the results shown in Table~\ref{table:data}. Corresponding values with energy-dependent freeze-out times had been obtained in Ref.\,\cite{fgw17}.\\

\begin{table}[!h]
\centering

\begin{tabular}{|l|l|l|l|}
\hline
  \multicolumn{1}{|c|}{System} & \multicolumn{1}{c|}{$\sqrt{s_{NN}}$ (\si{GeV})} & $A$ ($10^{24}$ \si{\per\second}) & $D$ ($10^{24}$ \si{\per\second}) \\ \hline
  PbPb                         & 17.2                               & 4.06 & 6.67 \\ \hline
  AuAu                         & 200                                  & 5.17 & 53.4 \\ \hline
\end{tabular}
\caption{Fitted values for the drift and diffusion coefficients in the case of normal \added{($q=1$)} diffusion\added{.}}
\label{table:data}
\end{table}

The failure to interpret the broad rapidity distributions observed in the stopping process of relativistic heavy-ion collisions within $q$-statistics refers specifically to the solution of the nonlinear Fokker-Planck equation Eq.\,(\ref{eq:3be}) which arises within nonextensive statistics \cite{tsa88,tsa96}. Among the abundant publications that compare $q$-statistics with data in particular for $p_\perp$-distributions
(e.g.\,\cite{wiwl09,ma15,wong15}), but also for rapidity distributions (\cite{ma15}), only few such as \cite{lav02,alb08} refer to an explicit -- but approximate -- solution of the basic nonlinear FPE. 
The result that was derived there makes use of the form of the stationary solution, replacing temperature and mean rapidity by time-dependent quantities. The outcome of this procedure does not agree with our numerical results for the explicit solution of the nonlinear FPE.
\section{Conclusion}
We have tested the nonextensive paradigm in a well-defined application to rapidity distributions in relativistic heavy-ion collisions.
For this problem a nonlinear Fokker-Planck equation is available which had been solved previously by Lavagno et al.~\cite{lav02,alb08,alb09}
using an approximate solution scheme. 

In our numerical solution we can reproduce neither the available data at SPS and RHIC energies for any value of $q \in (1,1.5)$, nor
the corresponding \deleted{approximate} solutions\added{ from Refs.~\cite{lav02,alb08,alb09}}.
The use of three different numerical methods with coinciding outcome and various cross checks 
with exact analytical results ensure the accuracy of \replaced{our}{the} numerical calculations.

This result casts doubt on the validity of the nonextensivity concept in statistical physics, which has often been applied
to interpret observables in relativistic heavy-ion collisions. Nevertheless, formulae derived from nonextensive statistics 
may still be used in phenomenological fits of transverse momentum distributions in relativistic collisions because
they allow to account for the observed transition from exponential to power law distributions that the $e^x_q$-function properly describes with
$q \in (1,1.5)$\added{. This transition had already earlier been modeled by Hagedorn in Ref.\cite{hag83} using an equivalent, but QCD-inspired formula.}



\replaced{The}{When comparing with} data for rapidity distributions in stopping and particle production \added{can be described using an}\deleted{we 
 start instead from the} unmodified Fokker-Planck equation with a linear diffusion term as in Boltzmann statistics\added{. Here we}\deleted{, and} determine \added{empirical} values of the diffusion coefficient that are necessary to reproduce the measured data, thus accounting phenomenologically
for both nonequilibrium thermal processes and collective expansion, without a nonlinear diffusion term.\\
\acknowledgments
We thank Andreas Mielke and Johannes H\"olck for discussions and remarks, and one of the referees for clarifying suggestions.

\bibliography{gw_17}

\end{document}